\documentclass[seceq,mbf,epsf,wrapft,amssymb]{ptptex}

\notypesetlogo                       

\title{
Quantum Field Theories in Nonextensive Tsallis Statistics
} 

\author{
Hiroaki {\sc Kohyama}\footnote{E-mail: kohyama@sci.osaka-cu.ac.jp} 
and 
Akira {\sc Ni\'egawa}\footnote{E-mail: niegawa@sci.osaka-cu.ac.jp}} 

\inst{Graduate School of Science, Osaka City University, Osaka 
558-8585, Japan} 

\abst{
Within the framework of Tsallis statistics with $q \simeq 1$, we 
construct a perturbation theory for treating relativistic 
quantum field systems. We find that there appear initial 
correlations, which do not exist in the Boltz\-mann-Gibbs statistics. 
Applying this framework to a quark-gluon plasma, we find that the so-called 
thermal masses of quarks and gluons are smaller than in the 
case of Boltzmann-Gibbs statistics.
}

\begin{document}

\maketitle

\section{Introduction} 
It is expected that a quark-gluon plasma will soon be 
produced in ultrarelativistic heavy-ion collision experiments at the 
BNL Relativistic Heavy Ion Collider (RHIC). The CERN Large Hadron 
Collider (LHC) will also soon be ready for experiments. A few 
observations regarding quark-gluon plasmas are in order. 
\begin{description}
\item{1)} According to hot QCD (statistical QCD at high temperature) 
\cite{leb}, in the chromoelectric sector of gluons in a quark-gluon 
plasma, a Debye screening mass develops, and as a consequence the 
chromoelectric-gluon exchange interaction is short range. By contrast,
 a Debye-like mass does not appear (at least to one-loop 
order) in the chromomagnetic sector of gluons, and therefore the 
chromomagnetic-gluon exchange interaction is long 
range.\footnote{It is worth mentioning here that, according to hot 
QED, the magnetic mass does not appear in the 
\lq\lq transverse-photon'' sector to all orders in perturbation 
theory \cite{gro}. Therefore, a magnetic-photon exchange 
interaction in an electron-positron-photon plasma is long 
range.}
\item{2)} The size of a region containing a 
quark-gluon plasma produced in the collision of heavy ions in RHIC 
experiments is not very large. In fact, the radius of heavy 
ions of mass number $\sim 200$ is $R \sim 7.9 \times 10^{- 15}$ m. 
At the highest RHIC energy, $\sqrt{s_{NN}} = 200$ GeV, the 
Lorentz-contraction factor is $\sim 1/100$, and therefore the longitudinal 
size of the system just after the collision is $L_l \sim 7.9 \times 
10^{- 17}$ m, and it increases due to the expansion 
of the system. Contrastingly, the chromoelectric Debye mass, 
given by \cite{leb} $m_D \simeq 1.2 g T$ (with $g$ the QCD coupling 
constant and $T$ the temperature of the plasma) for three quark 
flavors, is $\sim 2.8$ GeV for $T = 1$ GeV (where $g^2 / 4 \pi 
\simeq 0.43$), which corresponds to the Debye screening 
length\footnote{The quark counterpart \cite{leb} of the Debye 
screening length is much longer, specifically, $l_D \sim 2.1 \times 10^{- 
16}$ m for $T = 1$ GeV.} $l_D \sim 6.9 \times 10^{- 17}$ m. As 
mentioned above, the chromomagnetic \lq\lq Debye mass'', if it exists, is 
at most of $O (g^2 T)$. Thus the \lq\lq Debye screening length'' in 
the chromomagnetic sector is much larger than $l_D$. 
\item{3)} The single-particle transverse momentum distribution of 
hadrons, which are produced by heavy ion collisions, exhibits a 
power-law tail. 
\end{description}

Boltzmann-Gibbs statistics is valid for the systems that have the 
following properties: (i) compared to the size of the system, 
both the interactions and the memories are of short range, and (ii) 
the spacetime in which the system evolves is nonfractal. From the 
above observations 1) and 2), we find it doubtful that 
Boltzmann-Gibbs statistics can be applied in a strict sense 
to a quark-gluon plasma, at least in the early stages after a 
collision. It has been argued that the above observation 3) may be a 
sign of the fact that the object (a quark-gluon plasma or a 
hadronic fire ball) produced just after a heavy ion collision does 
not obey Boltzmann-Gibbs statistics (see, e.g., Ref. 3) and 
references therein). 

For a realistic description of a systems with long-range 
interactions, long-range memories and/or fractal structure, 
nonextensive generalization of Boltz\-mann-Gibbs statistics is 
essential. Such a generalization was proposed by Tsallis 
\cite{tsallis} seventeen years ago (so called Tsallis statistics). Since 
then, a variety of works have appeared concerning 
theoretical aspects of this form as well as its applications to 
various nonextensive systems. The most important problem among the 
former is to determine the extent to which Tsallis statistics is 
unique among other nonextensive statistics. A short review of this problem is given 
in \S2.1. Tsallis statistics has been successfully applied to a 
number of nonextensive systems. Examples\footnote{For a 
comprehensive list of references, see Ref.5). } 
are L\'evy-type anomalous diffusion \cite{ct}, Euler turbulence 
\cite{bm}, the specific heat of the hydrogen atom \cite{ls}, peculiar 
velocities in galaxies, \cite{al} and self-gravitating systems and 
related matters \cite{bm,iika}. Some theoretical frameworks in 
Boltzmann-Gibbs statistics have been generalized to the case of 
Tsallis statistics, e.g., linear response theory \cite{linear}, 
the Green function method \cite{green} and path integrals \cite{path}. 
Some approximation schemes, such as the ($1 - q$) expansion 
\cite{1q,lenn}, factorization approximation \cite{lenn,fac}, 
perturbation theory \cite{setsu} and the semi-classical expansion. 
\cite{semi} 

Tsallis statistics contains a (real) 
parameter $q$, which is a measure of nonextensiveness of the system, 
or more precisely, the range of interactions acting among the 
constituents of the system \cite{ante}. For short-range 
interactions, standard Boltzmann-Gibbs statistics is realized, 
which corresponds to Tsallis statistics with $q = 1$. 

With the above motivation, in this paper we treat
quantum field systems on the basis of Tsallis statistics 
assuming that $q$ differs from 1 by a very small amount. More 
precisely, our region of interest is that satisfying 
\begin{equation}
| 1 - q | \leq O (1 / (V T^3)^2)  \ll 1 , 
\label{region} 
\end{equation}
where $T$ is the \lq\lq temperature'' and $V$ is the volume of the 
system.\footnote{
The volume of a heavy ion with mass number $\sim 200$ is $V \simeq 
2.7 \times 10^{-4}$ $(\mbox{MeV}/\hbar c)^{- 3}$. (As mentioned 
above, at early stages after the collision, due to the Lorentz 
contraction, the volume of the produced quark-gluon plasma is much 
smaller than this $V$.) For $T = 200$, $500$ and $1000$ MeV, $V T^3 
\simeq 2.1 \times 10^3, 3.3 \times 10^4$ and $2.7 \times 10^5$, 
respectively.} (For $|1 - q|$ expansions in different contexts, 
see, e.g., Refs. 14) and 15), together with references 
therein.) 

In \S2, brief introductions to Tsallis statistics and the 
closed time-path (CTP) formalism \cite{leb,sch,chou} for treating 
quantum field systems are given. In \S3, we deduce the form for 
the CTP propagators in Tsallis statistics. In \S4, we discuss 
physical implications of the results and present a procedure for 
computing higher-order corrections to the propagators. 
\section{Preliminaries} 
\subsection{Tsallis statistics} 
Throughout this paper, we use units in which $k_B = \hbar = c = 1$. 
Tsallis postulates \cite{tsallis} the following form for the 
generalized entropy (the Tsallis entropy):
\begin{equation}
S_q [\hat{\rho}] = \frac{\left( 1 - \mbox{Tr} \hat{\rho}^q 
\right)}{q - 1} \, , 
\label{entro} 
\end{equation}
Here $q$ is a real-number parameter and $\hat{\rho}$ is the density 
operator (Tr $\hat{\rho} = 1$). In the limit $q \to 1$, Eq. 
(\ref{entro}) reduces to the standard Boltzmann-Gibbs-Shannon 
entropy $S_1 = - \mbox{Tr} \hat{\rho} \ln \hat{\rho}$. However, contrast 
to the case of Boltzmann-Gibbs statistics, the form for the entropy 
in the case we consider presently  
cannot be uniquely deduced. The Tsallis entropy (\ref{entro}) meets 
the requirement of \lq\lq concavity'', the 
requirement which any entropy should satisfy \cite{req}. Since the
introduction of the Tsallis entropy, 
dos Santos \cite{dos} has demonstrated the uniqueness of the 
form appearing in Eq. (\ref{entro}) assuming pseudo-additivity, $S_{A + 
B} = S_A + S_B + (1 - q) S_A S_B$ (where \lq $A$' and \lq $B$' represent 
the systems in question, and \lq $A + B$' represents the composite system of 
\lq $A$' and \lq $B$'), together with several other conditions. Hotta 
and Joichi \cite{hotta} have shown that Eq. (\ref{entro}) can be 
\lq\lq derived'' from less restrictive requirements, namely, the 
composability condition, $S_{A + B} = \Omega (S_A , S_B)$ (with 
$\Omega$ being some function), and the ansatz $S [\hat{\rho}] = C + 
\mbox{Tr} \phi (\hat{\rho})$ (with $C$ a constant and $\phi$ some 
function of $\hat{\rho}$), together with a few other conditions. 
Modified Tsallis entropies have also been proposed by several authors 
(see, e.g., Refs. 24) -- 27)). The connection between $S_q$ and the 
theory of quantum groups has been pointed out and discussed, e.g., 
in Refs. 26) and 28). 

The form of $\hat{\rho}$ is determined by generalizing the 
procedure employed in Boltzmann-Gibbs statistical mechanics. 
There had been some disagreement regarding the definition of the expectation value 
of an operator $\hat{A}$, but this issue has been settled, and it is now known
that this expectation value is given by the following \cite{tta}: 
\begin{equation}
\langle \hat{A} \rangle = \frac{\mbox{Tr} \hat{A} 
\hat{\rho}^q}{\mbox{Tr} \hat{\rho}^q} \, , 
\label{expect}
\end{equation}
which is called the $q$-expectation value and preserves various 
desirable properties. The form of the density operator $\hat{\rho}$ 
is determined by maximizing $S_q [\hat{\rho}]$ with the constraints 
$\mbox{Tr} \hat{\rho} = 1$ and $\langle \hat{H} \rangle = E$, where 
$\hat{H}$ is the Hamiltonian. Introducing Lagrange multipliers, we 
easily carry out the maximization and obtain 
\begin{eqnarray}
\hat{\rho} = Z_q^{- 1} \left[ 1 - (1 - q) \hat{H} / T \right]^{1 / 
(1 - q)} \, , \\  
Z_q = \mbox{Tr}' \left[ 1 - (1 - q) \hat{H} / T \right]^{1 / 
(1 - q)} \, . 
\label{mitu}
\end{eqnarray}
Here, $\mbox{Tr}'$ means that $\mbox{Tr}$ is taken over the energy 
eigenstates with $1 - (1 - q) E / T \geq 0$ with $E$ the eigenvalue 
of $\hat{H}$. For $1 - q < 0$, this restriction obviously 
does not apply. As mentioned in \S1, $q = 1$ corresponds to 
Boltzmann-Gibbs statistics, $\hat{\rho} \propto e^{- \beta \hat{H}}$ 
with $T \equiv 1 / \beta$ the temperature. In the following, we 
simply refer to $T$ in Eq. (\ref{mitu}) as the temperature. (For a 
thorough discussion of the temperature of nonextensive systems, 
see, e.g., Ref. 30) and references therein.) 

Before moving on, for convinience, we rewrite the formula (\ref{expect}) for the $q$-expectation value 
as 
\begin{eqnarray} 
\langle \hat{A} \rangle &=& \mbox{Tr}' \hat{\rho}' \hat{A} \, , 
\;\;\;\;\;\; \hat{\rho}' = \frac{\hat{\sigma}^q}{\mbox{Tr}' 
\hat{\sigma}^q} \, , 
\label{kiku} \\ 
\hat{\sigma}^q &=& \left[ 1 - (1 - q) \hat{H} / T \right]^{q / (1 - 
q)} = \left[ 1 - \epsilon \hat{H} / T \right]^{(1 - \epsilon) / 
\epsilon} \;\;\;\;\;\;\; (\epsilon = 1 - q) \nonumber \\ 
& = & \left[ 1 - \tilde{\epsilon} \tilde{\beta} 
\hat{H} \right]^{1 / \tilde{\epsilon}} 
\label{rewr} \\ 
\tilde{\epsilon} & \equiv & \frac{\epsilon}{1 - \epsilon} \, , 
\;\;\;\;\;\;\;\;\;\; \tilde{\beta} \equiv (1 - \epsilon) / T \, . 
\label{rep}
\end{eqnarray} 
\subsection{Closed time-path formalism}
For treating quantum field systems, we employ the 
closed time-path (CTP) formalism \cite{leb,sch,chou}. In the 
single-time representation of the CTP formalism, every field becomes two fields: 
$\phi \to (\phi_1, \phi_2)$. Here, $\phi_i$ $(i = 1, 2)$ is 
called the type-$i$ field, and $\phi_1$ is called the physical field. Then, 
the $n$-point Green function consists of $2^n$ components. At the 
very end of calculation $\phi_1$ and $\phi_2$ are set equal. 

In the following, we employ the complex scalar field theory governed by 
the Lagrangian density ${\cal L} = - \phi^\dagger (\partial^2 + 
m^2) \phi + {\cal L}_{int}$ with ${\cal L}_{int} = - (\lambda / 4) 
(\phi^\dagger \phi)^2$. Generalization to other field theories 
is straightforword. We restrict our consideration to the case in which the 
density operator is electrically neutral. 
\subsubsection{Single-time representation} 
The $2 n$-point Green function, which consists of $2^{2 n}$ 
components, is defined by  
\begin{eqnarray} 
&& G_{i_1 \cdots i_n; j_1 \cdots j_n} (x_1, \cdots , x_n; y_1, \cdots , y_n) 
\nonumber \\ 
&& \mbox{\hspace*{6ex}} = i (-)^n \mbox{Tr} \left[ T_c \left( 
\prod_{l = 1}^n \phi^{(H)}_{i_l} (x_l) \prod_{m = 1}^n 
\phi^{(H) \dagger}_{j_m} (y_m) \right) \hat{\rho} \right] \, , 
\label{midori}
\end{eqnarray} 
where the operators $\phi^{(H)}$ and $\phi^{(H) \dagger}$ are the 
Heisenberg field operators, and $\hat{\rho}$ is the density operator. 
Note that, for Tsallis statistics, we have $\hat{\rho} = \hat{\rho}'$ 
[Eq. (\ref{kiku})]. Here, $T_c$ is the \lq\lq ordering'' operator with the 
following properties: i) move the type-2 fields to the left of the 
type-1 fields, ii) rearrange the operators $\phi_1^{(H)}$ according to a 
time-ordering ($T_c \to T$), and iii) rearrange the operators $\phi_2^{(H)}$ 
according to an anti-time-ordering ($T_c \to \bar{T}$). 

Here we summarize the Feynman rules for computing $G$ perturbatively. 
The rules are the same as in the vacuum theory, except that the (bare) 
propagators and vertices take the following forms. 
\begin{enumerate} 
\item Propagators (bare two-point functions) 
\begin{description}
\item{1a)} Two-point propagators: 
\begin{eqnarray}
i \Delta_{11} (x - y) &=& \langle \mbox{T} \left( \phi_1 (x) 
\phi^\dagger_1 (y) \right) \rangle 
\, \rule[-2.85mm]{.14mm}{6.5mm} 
\raisebox{-2.35mm}{\scriptsize{$\; \phi_1 \equiv \phi$}} 
 \, , \nonumber \\ 
i \Delta_{22} (x - y) &=& \langle \overline{\mbox{T}} \left( 
\phi_2 (x) \phi^\dagger_2 (y) \right) \rangle 
\, \rule[-2.85mm]{.14mm}{6.5mm} 
\raisebox{-2.35mm}{\scriptsize{$\; \phi_2 \equiv \phi$}} 
\, , \nonumber \\ 
i \Delta_{12} (x - y) &=& \langle \phi^\dagger_2 (y) \phi_1 (x) 
\rangle 
\, \rule[-2.85mm]{.14mm}{6.5mm} 
\raisebox{-2.35mm}{\scriptsize{$\; \phi_1 = \phi_2 \equiv \phi$}} 
\, , \nonumber \\ 
i \Delta_{21} (x - y) &=& \langle \phi_2 (x) \phi^\dagger_1 (y) 
\rangle \, \rule[-2.85mm]{.14mm}{6.5mm} 
\raisebox{-2.35mm}{\scriptsize{$\; \phi_1 = \phi_2 \equiv \phi$}} 
\, , 
\label{prop}
\end{eqnarray}
where $\phi$ and $\phi^\dagger$ are the interaction-picture fields 
and $\langle \cdots \rangle \equiv \mbox{Tr} \cdots \, \hat{\rho}_0$ with 
$\hat{\rho}_0$ the \lq\lq bare'' density operator. 
\item{1b)} \lq\lq $2n$-point propagators'' ($2 \leq n$): The 
$2 n$-point propagators consist of $2^{2 n}$ components, all of 
which are the same: 
\begin{eqnarray}
&& {\cal C}_n (x_1, \cdots , x_n; y_1, \cdots , y_n) \, , \nonumber \\ 
&& \mbox{\hspace*{7ex}} \equiv \langle : \phi (x_1) \cdots \phi (x_n) 
\phi^\dagger (y_1) \cdots \phi^\dagger (y_n) : \rangle_c \;\;\;\;\;
\;\;\; , (2 \leq n) \, 
\label{mendoi} 
\end{eqnarray}
where $: \cdots :$ represents the operation of taking take a normal product and \lq $c$' stands 
for the contribution from the connected diagrams. ${\cal C}_n$ is 
called the initial correlation. 
\end{description}
\item Vertices \\ 
The vertex factor for a vertex at which type-1 fields meet is the 
same as in the vacuum theory, i.e. $- i \lambda$, while the vertex 
factor at which type-2 fields meet is $i \lambda$. 
\end{enumerate} 
\subsubsection{Physical representation} 
Let us introduce $\phi_c$ and $\phi_\Delta$ (and their hermitian 
conjugates) through the relations
\[
\phi_c = \frac{1}{2} (\phi_1 + \phi_2) \, , \;\;\;\;\;\;\;\; 
\phi_\Delta = \phi_1 - \phi_2 \, . 
\]
Using these relations, the single-time representation outlined above 
can be transformed into the representation written in terms of 
$\phi_c$, $\phi_\Delta$, $\phi_c^\dagger$ and $\phi_\Delta^\dagger$, 
which is called the physical representation. 

The $2 n$-point Green function is defined 
by\footnote{The normalization of $\tilde{G}$ here is different from that of
its counterpart in Ref. 21).} 
\begin{eqnarray} 
&& \tilde{G}_{c \cdots c \Delta \cdots \Delta; c \cdots c \Delta \cdots \Delta} 
(x_1, \cdots , x_n; y_1, \cdots , y_n) \nonumber \\ 
&& \mbox{\hspace*{2ex}} = i (-)^n \mbox{Tr} \left[ T_c \left( 
\prod_{i = 1}^{m_1} \phi^{(H)}_c (x_i) \prod_{j = m_1 + 1}^n 
\phi^{(H)}_\Delta (x_j) \prod_{k = 1}^{m_2} \phi^{(H) \dagger}_c 
(y_k) \prod_{l = m_2 + 1}^n \phi^{(H) \dagger}_\Delta (y_l) \right) 
\hat{\rho} \right] \, , \nonumber \\ 
&& 
\label{midorid}
\end{eqnarray} 

The propagators and vertices in the Feynman rules are as follows. 
\begin{enumerate} 
\item Propagators 
\begin{description}
\item{1a)} Two-point propagators: 
\begin{eqnarray}
i \tilde{\Delta}_{\Delta \Delta} &=& 0 \, , \nonumber \\ 
i \tilde{\Delta}_{cc} &=& i (\Delta_{11} + \Delta_{22}) / 2 \equiv i 
\Delta_c / 2 \, , \nonumber \\ 
i \tilde{\Delta}_{\Delta c} &=& i (\Delta_{11} - \Delta_{21}) \equiv 
i \Delta_A \, , \nonumber \\ 
i \tilde{\Delta}_{c \Delta} &=& i (\Delta_{11} - \Delta_{12}) \equiv 
i \Delta_R \, , 
\label{propd}
\end{eqnarray}
where $\Delta_{R (A)}$ is the retarded (advanced) propagator, which, 
in momentum space, reads 
\begin{equation}
\Delta_{R (A)} (P) = \frac{1}{P^2 - m^2 \pm i p_0 0^+} \, . 
\label{retadv}
\end{equation}
\item{1b)} Initial correlations: Among the $2^{2n}$ components of 
the $2 n$-point propagators $\tilde{\cal C}_n$ $(2 \leq n)$, only the $
(cc \cdots c; cc \cdots c)$-components are nonzero; 
\begin{eqnarray}
&& \left( \tilde{\cal C}_n \right)_{c \cdots c; c \cdots c} (x_1, \cdots , 
x_n; y_1, \cdots , y_n) = {\cal C}_n (x_1, \cdots , x_n; y_1, \cdots , y_n) 
\, , 
\label{mendoid}
\end{eqnarray}
where ${\cal C}_n$ is as in Eq. (\ref{mendoi}). 
\end{description}
\item Vertices \\ 
The vertex factors for the vertices $\phi_c^\dagger \phi_c^\dagger 
\phi_c \phi_\Delta$, $\phi_c^\dagger \phi_\Delta^\dagger \phi_c 
\phi_c$, $\phi_c^\dagger \phi_\Delta^\dagger \phi_\Delta 
\phi_\Delta$ and $\phi_\Delta^\dagger \phi_\Delta^\dagger \phi_c 
\phi_\Delta$ are $- i \lambda$, $- i \lambda$, $- i \lambda / 4$ 
and $- i \lambda / 4$, respectively. All other vertices vanish. 
\end{enumerate} 
\subsubsection{Forms of the propagators}
From this point, we restrict our attention to the case $\rho_0 = \rho_0 
(\hat{H}_0)$, with $\hat{H}_0$ the free Hamiltonian. Then, the system 
under consideration is spacetime-translation invariant, and therefore 
can go to momentum space. To obtain the forms of the 
propagators, we first construct single-particle wave functions by 
adopting a fixed volume ($V$) quantization with discrete momenta, 
${\bf p} = \left( 2 \pi / V^{1 / 3} \right) {\bf n}$, where ${\bf n} 
= (n_1, n_2, n_3),$ with $n_1, n_2, n_3$ integers. The complex scalar 
fields $\phi (x)$ are decomposed by using the plane-wave 
basis constructed in this manner, 
\begin{eqnarray}
\phi (x) &=& \sum_{\bf p} \frac{1}{(2 E_{\bf p} V)^{1 / 2}} \left[ 
a_{\bf p} e^{- i (E_{\bf p} x_0 - {\bf p} \cdot {\bf x})} + b_{\bf 
p}^\dagger e^{i (E_{\bf p} x_0 - {\bf p} \cdot {\bf x})} \right] 
\, , 
\label{ya} \\ 
&& \left[ a_{\bf p}, a_{{\bf p}'}^\dagger \right] = 
\left[ b_{\bf p}, b_{{\bf p}'}^\dagger \right] = \delta_{{\bf p}, 
{\bf p}'} \, , 
\label{comm}
\end{eqnarray}
where $E_{\bf p} = \sqrt{{\bf p}^2 + m^2}$ is the single-particle 
energy. Then, $\hat{H}_0$ becomes 
\begin{equation}
\hat{H}_0 = \sum_{\bf p} E_{\bf p} \left( a_{\bf p}^\dagger a_{\bf 
p} + b_{\bf p}^\dagger b_{\bf p} \right) \, . 
\label{hamil}
\end{equation}

The form of $\Delta_{i j} (P)$ is obtained by substituting Eq. 
(\ref{ya}) and its hermitian conjugate for $\phi$ and 
$\phi^\dagger$, respectively, into Eq. (\ref{prop}). Then, taking the 
large-volume limit, $V \to \infty$, we obtain (after taking the Fourier 
transform) 
\begin{eqnarray}
\Delta_{i j} (P) &=& \Delta_{i j}^{(0)} (P) + \Delta_\beta (P) \, , 
\nonumber \\ 
\Delta_{11}^{(0)} (P) &=& - \left( \Delta_{22}^{(0)} (P) \right)^* 
= \frac{1}{P^2 - m^2 + i 0^+} \, , \nonumber \\ 
\Delta_{12 (2 1)}^{(0)} (P) &=& - 2 \pi i \theta (\mp p_0) \delta 
(P^2 - m^2) \, , \nonumber \\ 
\Delta_\beta (P) &=& - 2 \pi i N (p_0) \delta (P^2 - m^2) \, , 
\label{denpa}
\end{eqnarray}
where $P^2 = p_0^2 - {\bf p}^2$ and $N (p_0)$ is the number-density 
function, 
\begin{eqnarray}
N (p_0) & = & \lim_{V \to \infty} \left[ \theta (p_0) \mbox{Tr} 
a_{\bf p}^\dagger a_{\bf p} \hat{\rho}_0 + \theta (- p_0) \mbox{Tr} 
b_{\bf p}^\dagger b_{\bf p} \hat{\rho}_0 \right] \nonumber \\ 
& \equiv & \mbox{Tr}  \hat{N} (p_0) \, . 
\label{ro} 
\end{eqnarray}

Similarly, we obtain the form for ${\cal C}_n$ [Eq. (\ref{mendoi})] as follows: 
\begin{eqnarray}
&& {\cal C}_n = \int \left( \prod_{i = 1}^n \frac{d^{\, 4} 
P_i}{(2 \pi)^4} \, 2 \pi \delta (P^2_i) \theta (p_i^0) \right) 
\sum_{l = 0}^n \langle \prod_{i = 1}^l  \hat{N} (p_{i 0}) 
\prod_{j = l + 1}^n  \hat{N} (- p_{j 0}) \rangle_c \nonumber \\ 
&& \mbox{\hspace*{7ex}} \times \sum_{i_1 < \cdots < i_l} \; 
\sum_{j_1 < \cdots < j_l} \left[ \left\{ e^{- i P_1 \cdot (x_{i_1} 
- y_{j_1})} \cdots e^{- i P_l \cdot (x_{i_l} - y_{j_l})} + 
\mbox{perms.} \right\} \right. \nonumber \\ 
&& \mbox{\hspace*{7ex}} \left. \times \left\{ e^{i P_{l+1} \cdot 
(x_{i'_1} - y_{j'_1})} \cdots e^{ i P_n \cdot (x_{i'_{n - 
l}} - y_{j'_{n - l}})} + \mbox{perms.} \right\} \right] \, . 
\label{incorr} 
\end{eqnarray}
Here, $P \cdot (x - y) = p_0 (x_0 - y_0) - {\bf p} \cdot ({\bf x} - 
{\bf y})$ and $\sum_{i_1 < \cdots < i_l}$ ($\sum_{j_1 < \cdots < j_l}$) 
is the summation over all possible choices of $i_1 \cdots i_l (j_1 \cdots j_l)$ from among the 
values $1,2,\cdots n$,subject to the conditins $i_1 < i_2 < 
\cdots < i_l$ ($j_1 < j_2 < \cdots < j_l$). $i_1' < 
i_2' < \cdots < i_{n - l}'$ ($j_1' < j_2' < \cdots < j_{n - l}'$) is 
obtained from $1, 2, \cdots, n$ by removing $i_1 < i_2 < \cdots < i_l$ 
($j_1 < j_2 < \cdots < j_l$). The first \lq perms.' indicates that all 
permutations among $(j_1, j_2, \cdots, j_n) $ are taken and the second \lq perms.' 
indicates that all permutations among $(j_1', j_2', \cdots, j_{n - 
l}')$ are taken. 

For Tsallis statistics [cf. Eq. (\ref{kiku}) and (\ref{rewr})], we have 
\begin{equation}
\hat{\rho}_0 = \frac{\hat{\sigma}^q_0}{\mbox{Tr}' \hat{\sigma}_0^q} 
= \frac{\left[ 1 - \tilde{\epsilon} \tilde{\beta} 
\hat{H}_0 \right]^{1 / \tilde{\epsilon}}}{\mbox{Tr}' \left[ 1 - 
\tilde{\epsilon} \tilde{\beta} \hat{H}_0 \right]^{1 / 
\tilde{\epsilon}}} \, . 
\label{11d}
\end{equation}
This is invariant under charge conjugation, and therefore we have $N (p_0) = N 
(|p_0|)$. Equation (\ref{incorr}) is simplified as 
\begin{eqnarray}
&& {\cal C}_n = \int \left( \prod_{i = 1}^n \frac{d^{\, 4} 
P_i}{(2 \pi)^4} \, 2 \pi \delta (P^2_i) \right) \langle \prod_{i = 
1}^n  \hat{N} (|p_{i 0}|) \rangle_c \nonumber \\ 
&& \mbox{\hspace*{7ex}} \times \left[ e^{- i P_1 \cdot (x_1 
- y_1)} \cdots e^{- i P_n \cdot (x_n - y_n)} + \mbox{perms.} 
\right] \, , 
\label{ai} 
\end{eqnarray}
where \lq perms.' indicates that all permutations among $(y_1, y_2, 
\cdots, y_n)$ are taken. 
\subsubsection{Gibbs ensemble}
A standard Gibbs ensemble with temperature $T$ $(= 1 / \beta)$ and 
vanishing chemical potential is described by $\hat{\rho}_0 = e^{- 
\beta \hat{H}_0} / \mbox{Tr} e^{- \beta \hat{H}_0}$. The initial 
correlation ${\cal C}_n$ $(2 \leq n)$, given in Eq. (\ref{mendoi}), 
vanishes. 

From Eq. (\ref{hamil}) with Eq. (\ref{comm}), in the 
limit $V \to \infty$, we obtain   
\begin{eqnarray}
\mbox{Tr} e^{- \beta \hat{H}_0} & \equiv & e^{{\cal P}_0 \beta V} 
\nonumber \\ 
& = & \lim_{V \to \infty} \prod_{\bf p} \left( 1 - e^{- \beta E_{\bf 
p}} \right)^{- 2} \nonumber \\ &=& \exp \left[ - 2 \times V \int 
\frac{d^{\, 3} p}{(2 \pi)^3} \, \ln \left( 1 - e^{- \beta E_{\bf p}} 
\right) \right] \, , 
\label{A} \\ 
N (|p_0|) \, \rule[-2.85mm]{.14mm}{6.5mm} 
\raisebox{-2.35mm}{\scriptsize{$\; |p_0| = E_{\bf p}$}} &=& 
\lim_{V \to \infty} \frac{\mbox{Tr} a_{\bf p}^\dagger 
a_{\bf p} e^{- \beta \hat{H}_0}}{\mbox{Tr} e^{- \beta \hat{H}_0}} 
= \lim_{V \to \infty} \frac{\mbox{Tr} b_{\bf p}^\dagger b_{\bf p} 
e^{- \beta \hat{H}_0}}{\mbox{Tr} e^{- \beta \hat{H}_0}} \nonumber \\ 
& = & \frac{\left( e^{\beta E_{\bf p}} - 1 \right)^{- 1} \, e^{{\cal 
P}_0 \beta V}}{e^{{\cal P}_0 \beta V}} \nonumber \\ 
&=& \frac{1}{e^{\beta E_{\bf p}} - 1} \equiv 
N_{\mbox{\scriptsize{BE}}} (E_{\bf p}) \, . 
\label{press}
\end{eqnarray}
Here ${\cal P}_0$ is the pressure of the free complex scalar field 
system and $N_{\mbox{\scriptsize{BE}}}$ is the familiar Bose 
distribution function. The factor of $2$ in Eq. (\ref{A}) corresponds to 
the number of degrees of freedom of the complex scalar field. 
A straightforward manipulation of Eq. (\ref{A}) yields, for $m \beta 
<< 1$, 
\begin{eqnarray}
\mbox{Tr} e^{- \beta \hat{H}_0} &=& e^{2 \times 
C_{\mbox{\scriptsize{BE}}} V / (3 \beta^3)} \, , 
\label{tee1} \\ 
C_{\mbox{\scriptsize{BE}}} &=& \frac{\pi^2}{3 0} \left( 1 - 
\frac{15}{4 \pi^2} (m \beta)^2 + \cdots \right) \, . 
\label{tee}
\end{eqnarray}
More generally, for a system that consists of single or several 
kinds of bosons and/or fermions, we have 
\begin{equation}
\mbox{Tr} e^{- \beta \hat{H}_0} = \exp \left[ \left( \sum_i 
n_{df}^{(i)} C_{\mbox{\scriptsize{BE}}}^{(i)} + \sum_j n_{df}^{(j)} 
C_{\mbox{\scriptsize{FD}}}^{(j)} \right) V / (3 \beta^3) \right] \, 
, 
\label{ome}
\end{equation}
where $i$ and $j$ label the kinds of bosons and fermions, 
respectively. The quantity $n_{df}^{(i)}$ ($n_{df}^{(j)}$) is the number of degrees of 
freedom of the $i$th kind of boson ($j$th kind of fermion) and 
\begin{eqnarray}
C_{\mbox{\scriptsize{BE}}}^{(i)} & = & \frac{\pi^2}{3 0} \left( 1 - 
\frac{15}{4 \pi^2} (m_i \beta)^2 + \cdots \right) \, , \nonumber \\ 
C_{\mbox{\scriptsize{FD}}}^{(j)} & = & \frac{\pi^2}{3 0} \left( 
\frac{7}{8} - \frac{15}{8 \pi^2} (m_j \beta)^2 + \cdots \right) \, . 
\label{omef}
\end{eqnarray}
It is worth mentioning that when we replace $\mbox{Tr} \cdots$ in Eq. 
(\ref{A}) with $\mbox{Tr}'\cdots$ (for $\epsilon > 0$) [cf. Eq. 
(\ref{11d})], terms of $O (e^{- 1 / \epsilon} / \epsilon^3)$ and 
$O ((\beta m / \epsilon)^2 e^{- 1 / \epsilon})$ appear in Eqs. 
(\ref{tee}) and (\ref{omef}). Such terms can safely be ignored for 
$\epsilon \ll 1$ [cf. Eq. (\ref{region})]. 
\section{Computation of the propagators} 
In this section, we compute the propagators on the basis of 
Tsallis statistics. 
\subsection{Preliminaries} 
$\hat{\sigma}_0^q$, appearing in Eq. (\ref{11d}), is expanded as follows:
\begin{eqnarray}
\hat{\sigma}_0^q &=& \exp \left[ \frac{1}{\tilde{\epsilon}} 
\ln \left( 1 - \tilde{\epsilon} \tilde{\beta} \hat{H}_0 \right) 
\right] \\ 
&=& e^{- \tilde{\beta} \hat{H}_0} \exp \left[ - \sum_{l = 
1}^\infty \frac{\tilde{\epsilon}^l (\tilde{\beta} \hat{H}_0)^{l + 
1}}{l + 1} \right] \nonumber \\ 
&=& e^{- \tilde{\beta} \hat{H}_0} \prod_{l = 1}^\infty \left[ 
\sum_{n = 0}^\infty \frac{1}{n !} \left( - \frac{\tilde{\epsilon}^l 
(\tilde{\beta} \hat{H}_0)^{l + 1}}{l + 1} \right)^n \right] \, . 
\label{saishuu} 
\end{eqnarray}
The region of interest is that defined by Eq. (\ref{region}). In the 
following, we compute field propagators up to and including 
$O (\epsilon)$ terms, with $\epsilon \leq O (1 / (V T^3)^2)$. For 
this purpose, for $\hat{\sigma}_0^q$, we should employ the form 
\begin{eqnarray} 
\hat{\sigma}_0^q &=& \sum_{l = 0}^\infty \frac{1}{l !} \left( - 
\frac{\tilde{\epsilon}}{2} \right)^l \left[ \left( \tilde{\beta} 
\hat{H}_0 \right)^{2 l} - \frac{\tilde{\epsilon}^2}{3} \left( 
\tilde{\beta} \hat{H}_0 \right)^{2 l + 3} 
+ \frac{\tilde{\epsilon}^4}{18} \left( \tilde{\beta} 
\hat{H}_0 \right)^{2 l + 6} \right. \nonumber \\ 
&& \left. - \frac{\tilde{\epsilon}^3}{4} \left( \tilde{\beta} 
\hat{H}_0 \right)^{2 l + 4} + \cdots \right] e^{- \tilde{\beta} 
\hat{H}_0} \nonumber \\ 
&=& \sum_{l = 0}^\infty \frac{1}{l !} \left( - 
\frac{\tilde{\epsilon}}{2} \right)^l \tilde{\beta}^{2 l} \left[ 
\partial^{2 l}_{\tilde{\beta}} + \frac{\tilde{\epsilon}^2}{3} 
\tilde{\beta}^3 \partial^{2 l + 3}_{\tilde{\beta}} + 
\frac{\tilde{\epsilon}^4}{18} \tilde{\beta}^6 \partial^{2 l + 
6}_{\tilde{\beta}} - \frac{\tilde{\epsilon}^3}{4} \tilde{\beta}^4 
\partial^{2 l + 4}_{\tilde{\beta}} + \cdots \right] e^{- \tilde{\beta} 
\hat{H}_0} \nonumber \\ 
& \equiv & {\cal D}_{\tilde{\beta}} e^{- \tilde{\beta} \hat{H}_0} 
\label{meii} 
\end{eqnarray} 
and for $\beta^N \partial^N \left( \mbox{Tr} e^{- \beta \hat{H}_0} 
\right) / \partial \beta^N$, the form
\begin{eqnarray} 
&& \beta^N \frac{\partial^N}{\partial \beta^N} \mbox{Tr} e^{- \beta 
\hat{H}_0} = \beta^N \frac{\partial^N}{\partial \beta^N} e^{2 
C_{\mbox{\scriptsize{BE}}} V / (3 \beta^3)} \nonumber \\ 
& & \mbox{\hspace*{3ex}} = \frac{(-)^N}{\beta^{3 N}} \biggl[ (2 
C_{\mbox{\scriptsize{BE}}} V)^N + 2 N (N- 1) (2 
C_{\mbox{\scriptsize{BE}}} V)^{N - 1} \beta^3  \nonumber \\ 
&& \mbox{\hspace*{6ex}}  + \frac{2 N (N - 1) (N - 2) (3 N - 
4) (2 C_{\mbox{\scriptsize{BE}}} V)^{N - 2}}{3} \beta^6 + \cdots 
\biggr] e^{2 C_{\mbox{\scriptsize{BE}}} V / (3 \beta^3)} \, . 
\nonumber \\ 
&& 
\label{form} 
\end{eqnarray} 
Here, the terms represented by \lq\lq $\cdots$'' yield at most $O (\epsilon^{3/ 2})$ 
contribution to the propagators. 

In the reminder of the paper, for simplicity, we restrict ourselves to 
the massless case, $m = 0$, and therefore we have $C_{\mbox{\scriptsize{BE}}} = 
\pi^2 / 30$ and $E_{\bf p} = |{\bf p}| = p$. Generalization to the 
massive case is straightforward. 

Equation (\ref{meii}) and the remark 
made at the end of \S2 guarantee that $\mbox{Tr}' \hat{\sigma}_0^q \simeq 
\mbox{Tr} \hat{\sigma}_0^q$. Then, using Eqs. (\ref{meii}), (\ref{tee1}) 
and (\ref{form}), we get
\begin{eqnarray} 
\mbox{Tr} \hat{\sigma}_0^q & = & {\cal D}_{\tilde{\beta}} \mbox{Tr} 
e^{- \tilde{\beta} \hat{H}_0} \nonumber \\ 
&=& \sum_{l = 0}^\infty \frac{1}{l !} (- \tilde{y})^l \left[ 1 + 4 
\{ 2 l (l - 1) + l \} \left( \frac{\tilde{\epsilon}}{2 \tilde{y}} 
\right)^{1 / 2} \right. \nonumber \\ 
&& + \frac{8}{3} l (l - 1) \{ 6 (l - 2) (l - 3) + 23 (l - 2) + 12 
\} \frac{\tilde{\epsilon}}{\tilde{y}} \nonumber \\ 
&& \left. - \frac{4 \tilde{y}}{3} \left( \frac{\tilde{\epsilon} 
\tilde{y}}{2}\right)^{1 / 2} - \frac{ 8 \tilde{y}}{3} \{ 2 l (l - 1) 
+ 7 l + 3 \} \tilde{\epsilon} + \frac{4 \tilde{y}^3}{9} 
\tilde{\epsilon} - \tilde{y}^2 \tilde{\epsilon} + \cdots \right] e^{2 
C_{\mbox{\scriptsize{BE}}} V / (3 \tilde{\beta}^2)} \nonumber \\ 
& = & \left[ 1 + \frac{2}{3} \sqrt{2 
\tilde{y} \tilde{\epsilon}} (5 \tilde{y} - 3) + \tilde{\epsilon} 
\tilde{y} \left( \frac{100}{9} \tilde{y}^2 - \frac{131}{3} \tilde{y} 
+ 24 \right) + \cdots \right] e ^{- \tilde{y}} e^{2 
C_{\mbox{\scriptsize{BE}}} V / (3 \tilde{\beta}^2)} \, , \nonumber 
\\ 
&& 
\label{1kkai} 
\end{eqnarray} 
where 
\begin{equation}
\tilde{y} \equiv 2 C^2_{\mbox{\scriptsize{BE}}} \tilde{\epsilon} 
\left( V \tilde{T}^3 \right)^2. \;\;\;\;\;\;\;\; (\tilde{T} = 1 / 
\tilde{\beta})  
\label{wai} 
\end{equation}
\subsection{Two-point propagator $\hat{\Delta}_{i, j}$ $(i, j = 1, 
2)$} 
The propagator $\Delta_{i j} (P)$ is given by Eq. (\ref{denpa}), with $N (|p_0|) = N 
(p) \, \rule[-2.85mm]{.14mm}{6.5mm} 
\raisebox{-2.35mm}{\scriptsize{$\; p = |p_0|$}}$ for the 
number-density function $N (p_0)$. Then, from Eqs. (\ref{A}) and 
(\ref{press}), we have 
\begin{eqnarray} 
\lim_{V \to \infty} \mbox{Tr} a_{\bf p}^\dagger a_{\bf p} e^{- 
\beta \hat{H}_0} & = & \lim_{V \to \infty} \mbox{Tr} b_{\bf 
p}^\dagger b_{\bf p} e^{- \beta \hat{H}_0} \nonumber \\ 
& = & \frac{1}{e^{\beta p} - 1} e^{2 C_{\mbox{\scriptsize{BE}}} V / 
(3 \beta^3)} = N_{\mbox{\scriptsize{BE}}} (p) e^{2 
C_{\mbox{\scriptsize{BE}}} V / (3 \beta^3)} \, . 
\label{BG}
\end{eqnarray} 
Using Eqs. (\ref{meii}), (\ref{tee1}) with $m = 0$ and (\ref{BG}), 
we obtain 
\begin{eqnarray} 
&& \lim_{V \to \infty} \mbox{Tr} a_{\bf p}^\dagger a_{\bf p} 
\hat{\sigma}_0^q \nonumber \\ 
& & \mbox{\hspace*{3ex}} = {\cal D}_{\tilde{\beta}} \left( \tilde{N} 
e^{2 C_{\mbox{\scriptsize{BE}}} V / (3 \tilde{\beta}^3)} \right) 
\nonumber \\ 
& & \mbox{\hspace*{3ex}} = \tilde{N} \sum_{l = 0}^\infty 
\frac{1}{l !} \left( - \tilde{y} \right)^l \biggl[ 2 l \tilde{\beta} 
p (1 + \tilde{N}) \left( \frac{\tilde{\epsilon}}{2 \tilde{y}} 
\right)^{1 / 2}  \nonumber \\ 
&& \mbox{\hspace*{6ex}} + \frac{1}{2} \left\{ 8 l (l - 1) (2 l - 1) 
\tilde{\beta} p (1 + \tilde{N}) + l (2 l - 1) (\tilde{\beta} p)^2 (1 
+ \tilde{N}) (1 + 
2 \tilde{N}) \right\} \frac{\tilde{\epsilon}}{\tilde{y}} \nonumber 
\\ 
&& \mbox{\hspace*{6ex}}  - \frac{2}{3} (2 l + 3) \tilde{\beta} 
p (1 + \tilde{N}) \tilde{\epsilon} \tilde{y} + \cdots \biggr] e^{2 
C_{\mbox{\scriptsize{BE}}} V / (3 \tilde{\beta}^2)} + \tilde{N} 
\mbox{Tr} \hat{\sigma}^q_0 \nonumber \\ 
&& \mbox{\hspace*{3ex}} = \tilde{N} e^{- \tilde{y}} \biggl[ - 2 
\tilde{\beta} p (1 + \tilde{N}) \left( \frac{\tilde{\epsilon} 
\tilde{y}}{2} \right)^{1 / 2} + 4 \tilde{\beta} p (1 + \tilde{N}) 
\tilde{y} (3 - 2 \tilde{y}) \tilde{\epsilon}  \nonumber \\ 
&& \mbox{\hspace*{6ex}}  + \frac{(\tilde{\beta} p)^2}{2} (2 
\tilde{N}^2 + 3 \tilde{N} + 1) (2 \tilde{y} - 1) \tilde{\epsilon} - 
\frac{2}{3} \tilde{\beta} p (1 + \tilde{N}) \tilde{y} (3 - 2 
\tilde{y}) \tilde{\epsilon} + \cdots \biggr] 
e^{2 C_{\mbox{\scriptsize{BE}}} V / (3 \tilde{\beta}^3)} \nonumber 
\\ 
&& \mbox{\hspace*{6ex}} + \tilde{N} \mbox{Tr} \hat{\sigma}^q_0 \, , 
\label{2kai} 
\end{eqnarray} 
where $\tilde{N} = 1 / (e^{\tilde{\beta} p} - 1)$. From Eqs. 
(\ref{1kkai}) and (\ref{2kai}), we finally obtain for the 
number-density function, 
\begin{eqnarray} 
N (p) &=& \lim_{V \to \infty} \frac{\mbox{Tr} a_{\bf 
p}^\dagger a_{\bf p} \hat{\sigma}_0^q}{\mbox{Tr} \hat{\sigma}_0^q} 
\nonumber \\ 
&=& \tilde{N} \biggl[ 1 - \sqrt{2 \tilde{y} \tilde{\epsilon}} 
\tilde{\beta} p (1 + \tilde{N})  \nonumber \\ 
&&  + \tilde{\epsilon} \biggl\{ 6 \tilde{\beta} p (1 + 
\tilde{N}) \tilde{y} + \frac{(\tilde{\beta} p)^2}{2} (2 \tilde{N}^2 
+ 3 \tilde{N} + 1) (2 \tilde{y} - 1) \biggr\} + \cdots 
\biggr] \nonumber \\ 
&=& N_{\mbox{\scriptsize{BE}}} \biggl[ 1 - \sqrt{2 y \epsilon} \, 
\beta p (1 + N_{\mbox{\scriptsize{BE}}})  \nonumber \\ 
&& + \epsilon \biggl\{ (1 + 6 y) \beta p (1 + 
N_{\mbox{\scriptsize{BE}}}) + \frac{2 y - 1}{2} (\beta p)^2 
\left( 1 + N_{\mbox{\scriptsize{BE}}} \right) \left( 1 + 2 
N_{\mbox{\scriptsize{BE}}} \right) \biggr\} \nonumber \\ 
&&  + O (\epsilon^{3 / 2}) \biggr] \, , 
\label{fine} 
\end{eqnarray} 
where use has been made of Eq. (\ref{rep}). In Eq. (\ref{fine}), we have used
$N_{\mbox{\scriptsize{BE}}} = 1 / (e^{\beta p} - 1)$ and 
\begin{equation}
y = \frac{\pi^4}{450} \, \epsilon \, (V T^3)^2 \, . 
\label{yai}
\end{equation}

Generalization to other fields is straightforward. For a system of 
bosons and/or fermions, we have, for the number-density function,  
\begin{eqnarray} 
N_{\mbox{\scriptsize{b/f}}} (p) &=& N_{\mbox{\scriptsize{BE/FD}}} 
\biggl[ 1 - \sqrt{2 y \epsilon} \beta p (1 \pm 
N_{\mbox{\scriptsize{BE/FD}}}) 
+ \epsilon \biggl\{ (1 + 6 y) \beta p (1 \pm 
N_{\mbox{\scriptsize{BE/FD}}}) \nonumber \\ 
&&  + \frac{2 y - 1}{2} (\beta p)^2 (1 \pm 
N_{\mbox{\scriptsize{BE/FD}}} ) (1 \pm 2 
N_{\mbox{\scriptsize{BE/FD}}}) \biggr\} + O (\epsilon^{3 / 2}) 
\biggr] \, , 
\label{fineb} \\ 
y & = & \frac{\left( \sum_i n_{df}^{(i)} + (7 / 8) \sum_j 
n_{df}^{(j)} \right)^2 \pi^4}{1800} \, \epsilon \left( V 
T^3 \right)^2 . \;\; \left( \leq O (1) \right) \, 
\label{cbe} 
\end{eqnarray} 
Here \lq b' (\lq f') indicates a boson (fermion), 
$C_{\mbox{\scriptsize{BE/FD}}}^{(i) / (j)}$ is as in Eq. 
(\ref{omef}), with $m = 0$, and $N_{\mbox{\scriptsize{FD}}} = 1 / 
(e^{\beta p} + 1)$. 

Equation (\ref{fineb}) is valid for $|y| \leq O (1)$. We note that, from 
Eq. (\ref{cbe}), the quantities $y$ $(\propto \epsilon)$ and $\epsilon$ are of the 
same sign. Then, from Eq. (\ref{fineb}), we see that the $O 
(\sqrt{\epsilon})$ term is negative and invariant under $\epsilon \to - 
\epsilon$, so that  
$N_{\mbox{\scriptsize{BE/FD}}}-N_{\mbox{\scriptsize{b/f}}}$ 
is the same for $\epsilon = |\epsilon|$ as for $\epsilon = - |\epsilon|$
. The difference between the cases $\epsilon > 0$ and $\epsilon < 0$ 
arises at $O (\epsilon)$. 
\subsubsection*{Low-temperature limit: $\beta \to 
\infty$} 
In the limit $\beta p >> 1$, we have $N_{\mbox{\scriptsize{BE/FD}}} 
\sim e^{- \beta p}$, and then 
\[ 
N_{\mbox{\scriptsize{b/f}}} (p) 
=
\left[ 1 - \sqrt{2 y \epsilon} \, \beta p + 
(1 + 6 y) \epsilon \beta p 
+ \frac{1}{2}  (2 
y - 1)\epsilon(\beta p)^2 + O (\epsilon^{3 / 2}) \right] e^{- \beta p} \, . 
\]
The series in the square brackets on the R.H.S. here seems to be 
nonconvergent for 
\begin{equation}
1 / \sqrt{\epsilon} \lesssim \beta p \;\; \mbox{or} \;\; 1 / 
\sqrt{\epsilon y} \lesssim \beta p \, . 
\label{kiken}
\end{equation}
However, due to the factor $e^{- \beta p}$, the region (\ref{kiken}) 
is unimportant. We see that $[ N_{\mbox{\scriptsize{b/f}}} 
(p)]_{\epsilon > 0} < [ N_{\mbox{\scriptsize{b/f}}} (p) 
]_{\epsilon = - |\epsilon|}$. 
\subsubsection*{High-temperature limit: $\beta \to 
0$} 
In the limiting case $\beta p \simeq 0$, we have $N_{\mbox{\scriptsize{BE}}} 
\sim 1 / (\beta p)$ and $N_{\mbox{\scriptsize{FD}}} \simeq 1 / 2 - 
\beta p / 4$. Then, we obtain  
\begin{eqnarray}
N_{\mbox{\scriptsize{b}}} (p) &=& \frac{1}{\beta p} \left[ 1 - 
\sqrt{2 y \epsilon} + 8 y \epsilon + O (\epsilon^{3 / 2}) \right] 
\, , \nonumber \\ 
N_{\mbox{\scriptsize{f}}} (p) &=& \frac{1}{2} - \frac{1}{4} 
\beta p \left[ 1 + \sqrt{2 y \epsilon} - (1 + 6 y) \epsilon + O 
(\epsilon^{3 / 2}) \right] \, . 
\end{eqnarray}
In this case, we see that $[ N_{\mbox{\scriptsize{b/f}}} 
(p)]_{\epsilon > 0} > [ N_{\mbox{\scriptsize{b/f}}} (p) 
]_{\epsilon = - |\epsilon|}$. 
\subsection{Initial correlations ($2 n$-point propagators) 
${\cal C}_n$}
Here, we compute $\langle \prod_{i = 1}^n \hat{N} (|p_{i 0}|) 
\rangle_c$ $(|p_{i 0}| = p_i)$, appearing in Eq. (\ref{ai}), in the 
content of complex scalar field theory. A procedure similar to that above 
leading to Eq. (\ref{fineb}) in the present case yields [cf. Eq. (\ref{BG})] 
\begin{eqnarray}
\langle \prod_{i = 1}^n \hat{N} (p_i) \rangle & = & \frac{{\cal 
D}_{\tilde{\beta}} \left( \prod_{i = 1}^n \left( e^{\tilde{\beta} 
p_i} - 1 \right)^{- 1} e^{2 C_{\mbox{\scriptsize{BE}}} 
V/ (3 \tilde{\beta}^3)} \right)}{{\cal D}_{\tilde{\beta}} e^{2 
C_{\mbox{\scriptsize{BE}}} V/ (3 \tilde{\beta}^3)}} \nonumber \\ 
&=& \left( \prod_{i = 1}^n N_{\mbox{\scriptsize{BE}}} (p_i) \right) 
\biggl[ 1 - \sqrt{2 y \epsilon} \sum_{i = 1}^n \beta p_i (1 + 
N_{\mbox{\scriptsize{BE}}} (p_i))  \nonumber \\ 
&& + \epsilon \sum_{i = 1}^n \biggl\{ (1 + 6 y) \beta p_i (1 + 
N_{\mbox{\scriptsize{BE}}} (p_i))  \nonumber \\ 
&& + \frac{2 y - 1}{2} (\beta p_i)^2 (1 + N_{\mbox{\scriptsize{BE}}} 
(p_i)) (1 + 2 N_{\mbox{\scriptsize{BE}}} (p_i)) \nonumber \\ 
&&   + \sum_{j ( \neq i)} (2 y - 1) \beta^2 p_i p_j (1 + 
N_{\mbox{\scriptsize{BE}}} (p_i)) (1 + N_{\mbox{\scriptsize{BE}}} 
(p_j)) \biggr\} + O (\epsilon^{3 / 2}) \biggr] \, . \nonumber \\ 
&& 
\label{finebb} 
\end{eqnarray}
We are now in a position to compute the connected contribution, 
$\langle \prod_{i = 1}^n \hat{N} (p_i) \rangle_c$. For the sake of 
generality, from this point, we consider a system of bosons and/or 
fermions. We start with $n = 2$, in which case we have 
\begin{eqnarray}
\langle \hat{N} (p_1) \hat{N} (p_2) \rangle_c &=& \langle \hat{N} 
(p_1) \hat{N} (p_2) \rangle - N_{\mbox{\scriptsize{b/f}}} (p_1) 
N_{\mbox{\scriptsize{b/f}}} (p_2) \, . 
\end{eqnarray}
Then, using Eqs. (\ref{fineb}) and (\ref{finebb}), we obtain 
\begin{eqnarray}
\langle \hat{N} (p_1) \hat{N} (p_2) \rangle_c & = & - \epsilon 
\beta^2 p_1 p_2 N_{\mbox{\scriptsize{BE/FD}}} (p_1) 
N_{\mbox{\scriptsize{BE/FD}}} (p_2) \left( 1 \pm 
N_{\mbox{\scriptsize{BE/FD}}} (p_1) \right) \nonumber \\ 
&& \times \left( 1 \pm N_{\mbox{\scriptsize{BE/FD}}} (p_2) 
\right) + O (\epsilon^{3 / 2})\, , 
\label{22}
\end{eqnarray}
which is of $O (\epsilon)$. For $n = 3$, we have 
\begin{eqnarray}
&& \langle \hat{N} (p_1) \hat{N} (p_2) \hat{N} (p_3) \rangle_c 
\nonumber \\ 
&& \mbox{\hspace*{5ex}} = \langle \hat{N} (p_1) \hat{N} (p_2) 
\hat{N} (p_3) \rangle - N_{\mbox{\scriptsize{b/f}}} (p_1) 
N_{\mbox{\scriptsize{b/f}}} (p_2) N_{\mbox{\scriptsize{b/f}}} 
(p_3) \nonumber \\ 
&& \mbox{\hspace*{8ex}} - \langle \hat{N} (p_1) \hat{N} (p_2) 
\rangle_c N_{\mbox{\scriptsize{b/f}}} (p_3) - \langle \hat{N} 
(p_2) \hat{N} (p_3) \rangle_c N_{\mbox{\scriptsize{b/f}}} 
(p_1) \nonumber \\ 
&& \mbox{\hspace*{8ex}} - \langle \hat{N} (p_1) \hat{N} (p_3) 
\rangle_c N_{\mbox{\scriptsize{b/f}}} (p_2) \nonumber \\ 
&& \mbox{\hspace*{5ex}} = O (\epsilon^{3/2}) \, . 
\end{eqnarray}
In a similar manner, for $4 \leq n$, we obtain  
\begin{eqnarray}
\langle \prod_{i = 1}^n \hat{N} (p_i) \rangle_c &=& \langle 
\prod_{i = 1}^n \hat{N} (p_i) \rangle - \prod_{i = 1}^n 
N_{\mbox{\scriptsize{b/f}}} (p_i) \nonumber \\ 
&& - \sum_{\mbox{\scriptsize{perm.}}} \langle \hat{N} (p_1) \hat{N} 
(p_2) \rangle_c \prod_{i = 3}^n N{\mbox{\scriptsize{b/f}}} 
(p_i) + O (\epsilon^{3 / 2}) \nonumber \\ 
&=& O (\epsilon^{3 / 2}) \, . 
\end{eqnarray}
Thus, we conclude that, to $O (\epsilon)$, 
${\cal C}_n = 0$ for $3 \leq n$. ${\cal C}_2$ is given by 
Eq. (\ref{ai}) with $n = 2$ and Eq. (\ref{22}), which is of 
$O (\epsilon)$. 

Finally, we point out that in the case of Boltzmann-Gibbs statistics, we have 
$\langle \prod_{i = 1}^n \hat{N} (p_i) \rangle = 
\prod_{i = 1}^n \langle \hat{N} (p_i) \rangle$, and hence ${\cal C}_n 
= 0$. 
\section{Physical implications}
\subsection{Hard thermal loops in hot QCD}
According to QCD at high temperature (hot QCD), among the formally 
higher-order amplitudes are those that are of the same order of 
magnitude as their lowest-order counterparts \cite{leb}. This is the 
case for classes of amplitudes whose external momenta are all soft, i.e.,
$P^\mu = O (g T)$ with $g$ the QCD coupling constant. The relevant 
diagrams are the one-loop diagrams with hard loop momenta 
$[Q_{\mbox{\scriptsize{loop}}} = O (T)]$, and for this reason, these are called 
hard thermal loops (HTL). Each HTL amplitude is proportional to the 
characteristic mass, the gluon thermal mass ($m_g$), or the quark 
thermal mass ($m_q$): 
\begin{eqnarray}
m^2_g & = & \frac{1}{2} g^2 T^2 (1 + N_f / 6) = \frac{3 g^2}{\pi^2} 
(1 + N_f / 6) {\cal I}_{\mbox{\scriptsize{BG}}} \, , \nonumber \\ 
m_q^2 & = & \frac{1}{6} g^2 T^2 = \frac{g^2}{\pi^2} {\cal 
I}_{\mbox{\scriptsize{BG}}} \, , 
\label{mass}
\end{eqnarray}
where $N_f$ is the number of active (massless) quark flavors. In Eq. 
(\ref{mass}), we have introduced ${\cal I}_{\mbox{\scriptsize{BG}}}$ 
(where \lq BG' means \lq Boltzmann-Gibbs'), which is defined by 
\begin{equation}
{\cal I}_{\mbox{\scriptsize{BG}}} \equiv \int_0^\infty d p \, p 
N_{\mbox{\scriptsize{BE}}} (p) = 2 \int_0^\infty d p \, p 
N_{\mbox{\scriptsize{FD}}} (p) =  \pi^2 T^2 / 6 \, . 
\label{reff}
\end{equation}

For Tsallis statistics, the above ${\cal 
I}_{\mbox{\scriptsize{BG}}}$ is modified. From Eq. (\ref{fineb}), we 
can compute the Tsallis statistics counterpart of Eq. (\ref{reff}), 
${\cal I}_{\mbox{\scriptsize{T}}}$: 
\begin{eqnarray}
{\cal I}_{\mbox{\scriptsize{T}}} & \equiv & \int_0^\infty d p \, 
p N_{\mbox{\scriptsize{b}}} (p) = 2 \int_0^\infty d p \, p 
N_{\mbox{\scriptsize{f}}} (p) \nonumber \\ 
& = & \left[ 1 - 2 \sqrt{2 y \epsilon} + \epsilon (18 y - 1) + O 
(\epsilon^{3 / 2}) \right] {\cal I}_{\mbox{\scriptsize{BG}}} \, . 
\end{eqnarray}
Thus, we obtain for the gluon and quark thermal masses in 
Tsallis statistics, 
\begin{eqnarray}
\frac{ \left( m^2_{g / q} \right)_{\mbox{\scriptsize{T}}}}{\left( 
m^2_{g / q} \right)_{\mbox{\scriptsize{BG}}}} & = & 1 - 2 \sqrt{2 y 
\epsilon} + \epsilon (18 y - 1) + O (\epsilon^{3 / 2}) \, , 
\end{eqnarray}
where $\bigl( m^2_{g / q} \bigr)_{\mbox{\scriptsize{BG}}}$ is as 
in Eq. (\ref{mass}) and $y$ is as in Eq. (\ref{cbe}), with $n_{df} = 
16$ and $n_{df'} = 12 N_f$. We find that, to $O (\sqrt{\epsilon})$, 
$\bigl( m^2_{g / q} \bigr)_{\mbox{\scriptsize{T}}} < 
\bigl( m^2_{g / q} \bigr)_{\mbox{\scriptsize{BG}}}$. 
\subsection{Manifestation of the initial correlations} 
Here we give some examples of quantities in which ${\cal C}_n$ 
participates. We consider a complex scalar field system. 
(Generalization to other field systems is straightforward.) First, 
we note that each of the average values 
$\langle \phi_c \phi_c^\dagger \rangle$ $(\sim i 
\tilde{\Delta}_{cc})$, $\langle \phi_c \phi_c \phi_c^\dagger 
\phi_c^\dagger \rangle$ $(\sim i \tilde{\Delta}_{cccc})$, $\cdots$ that we 
know represents information that we possess concerning the statistical properties 
of the system.  

The linear response theory is formulated in terms of the so-called 
linear response function of the $n$-point correlation with the 
external field that couples to the quantity under consideration. 
The linear response functions are given by the appropriate Green 
functions, $\tilde{G}$, in the CTP formalism\cite{chou}. As an 
example, we take the field $\phi^\dagger$ ($\phi$) itself as such a 
quantity. Then, the linear response function of the $(n-1)$-point 
correlation $(n = 2, 4, \cdots)$ is given by $\tilde{G}_{c \cdots c; c \cdots 
c \Delta}$ ($\tilde{G}_{c \cdots \Delta; c \cdots c})$. Similarly, the 
nonlinear response functions are also given by the appropriate Green 
functions in the CTP formalism. For example, the second-order 
response functions of the two-point correlations are given by 
$\tilde{G}_{cc; \Delta \Delta}$, $\tilde{G}_{c \Delta; c \Delta}$ 
and $\tilde{G}_{\Delta \Delta; c c}$. 

The initial correlations, given in Eq. (\ref{mendoid}), contribute to the 
above-mentioned response functions as well as to the average 
correlations $\tilde{G}_{c \cdots; c \cdots c}$, and hence they can be 
\lq\lq measured'' in principle. For example, the diagrams that 
include one vertex and one four-point propagator ${\cal C}_2$ 
contribute to the four-point Green functions $\tilde{G}_{cc; cc}$, 
$\tilde{G}_{cc; c \Delta}$ and $\tilde{G}_{c \Delta; cc}$. Explicit 
computation of the contribution to, e.g., $\tilde{G}_{c c; c \Delta}$, 
using the form for ${\cal C}_2$ obtained in \S3.3 yields 
\begin{eqnarray}
&& \tilde{G}_{c c; c \Delta} (x_1, x_2; y_1, y_2) \nonumber \\ 
&& \mbox{\hspace*{7ex}} = \frac{\lambda}{2} \int d^{\, 4} z \left[ 
\left\{ 2 {\cal C}_2 (z, x_1; z, y_1) S_R (z - y_2) S_R (x_2 - z) + 
(x_1 \leftrightarrow x_2) \right\} \right. \nonumber \\ 
&& \mbox{\hspace*{11ex}} \left. + {\cal C}_2 (x_1, x_2; z, z) S_R 
(z - y_2) S_R (y_1 - z) \right] \nonumber \\ 
&& \mbox{\hspace*{7ex}} = \frac{\lambda}{2} \int \frac{d^{\, 4} 
P}{(2 \pi)^4} \int \frac{d^{\, 4} P_1}{(2 \pi)^4} \int \frac{d^{\, 
4} P_2}{(2 \pi)^4} 2 \pi \delta (P_1^2) 2 \pi \delta (P_2^2) \left\{ 
\left( N (p_1) N (p_2) \right)_T \right\}_c \nonumber \\ 
&& \mbox{\hspace*{11ex}} \times 
\biggl[ \left\{ 2 e^{- i [P \cdot (x_2 - y_2) + P_2 \cdot (x_1 - 
y_1)]} \left( S_R (P) \right)^2  \right. \nonumber \\ 
&& \mbox{\hspace*{15ex}} + 2 e^{- i [P \cdot (x_2 - y_2) + P_1 \cdot 
(x_2 - y_1) - P_2 \cdot (x_2 - x_1)]} S_R (P) S_R (P + P_1 - P_2) 
\nonumber \\ 
&& \left. \mbox{\hspace*{15ex}} + e^{- i [P \cdot (y_1 - y_2) + P_1 
\cdot (x_1 - y_1) + P_2 \cdot (x_2 - y_1)]} S_R (P) S_R (P - P_1 - 
P_2) \right\} \nonumber \\ 
&&  \mbox{\hspace*{15ex}} + \left\{ x_1 \leftrightarrow x_2 
\right\} \biggr] \, . 
\label{tui}
\end{eqnarray}
Note that $\tilde{G}_{cc; c \Delta} (x_1, x_2 ; y_1, y_2)$ can also 
be rewritten in the form \cite{chou} 
\begin{eqnarray}
&& \tilde{G}_{c c ; c \Delta} (x_1, x_2; y_1, y_2) \nonumber \\ 
&& \mbox{\hspace*{7ex}} = \frac{i}{4} 
\sum_{\mbox{\scriptsize{perms.}}} \left[ \theta (x_1, x_2, y_1, y_2) 
\langle \left[ \left\{ \left\{ \phi (x_1), \phi (x_2) \right\}, 
\phi^\dagger (y_1) \right\}, \phi^\dagger (y_2) \right] \rangle 
\right. \nonumber \\ 
&& \mbox{\hspace*{7ex}} + \theta (x_1, x_2, y_2, 
y_1) \langle \left\{ \left[ \left\{ \phi (x_1), \phi (x_2) \right\}, 
\phi^\dagger (y_2) \right], \phi^\dagger (y_1) \right\} \rangle 
\nonumber \\ 
&& \mbox{\hspace*{11ex}} \left. + \theta (x_1, y_2, x_2, y_1) 
\langle \left\{ \left\{ \left[ \phi (x_1), \phi^\dagger (y_2) 
\right], \phi (x_2) \right\}, \phi^\dagger (y_1) \right\} \rangle 
\right] \, , 
\end{eqnarray}
where \lq perms.' indicates that all permutations among $(x_1, x_2, 
y_1)$ are taken, and $\theta (x, y, u, v) \equiv \theta (x_0 - y_0) \theta 
(y_0 - u_0) \theta (u_0 - v_0)$. From this formula, we see that 
$\tilde{G}_{c c ; c \Delta} = 0$ for $y_{20} > x_{1 0}, x_{2 0}, 
y_{1 0}$, as should be the case. As must be the case, $\tilde{G}_{cc; c 
\Delta}$ in Eq. (\ref{tui}) satisfies this property. 
\subsection{Remarks on higher-order corrections} 
Higher-order corrections to the CTP propagators come from the 
following two sources. 
\begin{enumerate}
\item Corrections due to the interactions, which, as in the 
case of Boltzmann-Gibbs statistics, are obtained by computing the 
vertex-insertion diagrams. 
\item Corrections arising from the corrections to the pressure 
${\cal P}$ computed within the Boltzmann-Gibbs statistics [cf. Eqs. 
(\ref{A}) and (\ref{press})]. For hot QCD, the corrections to 
${\cal P}$ are known \cite{kap} up to $O [g^5 \ln (1 / g)]$. 
\end{enumerate}

\section*{Acknowledgements}
The authors are grateful for useful discussions at the Workshop on 
Thermal Field Theories and their Applications, held August 9 - 11, 
2004, at the Yukawa 
Institute for Theoretical Physics, Kyoto, Japan.
A. N. is supported in part by a Grant-in-Aid for Scientific 
Research [(C)(2) No. 17540271] from the Ministry of Education, 
Culture, Sports, Science and Technology, Japan. 

\end{document}